\documentclass[twocolumn,aps,pra,superscriptaddress,showpacs]{revtex4}
\usepackage{graphicx,dcolumn}
\usepackage{amsfonts,amssymb,amsmath,bbm}
\usepackage[english]{babel}
\begin{document}

\title{Nonlinear coherent transport of waves in disordered media}
\author{Thomas Wellens}
%\email{Thomas.Wellens@physik.uni-erlangen.de}
\affiliation{Institut f\"ur Theoretische Physik, Universit\"at
  Erlangen-N\"urnberg, Staudtstr. 7, 91058 Erlangen, Germany}
\author{Beno\^it Gr\'emaud}
\affiliation{Laboratoire Kastler Brossel, Ecole Normale Sup\'erieure,
  CNRS, UPMC,
4 place Jussieu, 75252 Paris Cedex 05, France}

\date{\today}
\begin{abstract}
We present a diagrammatic theory for coherent
backscattering from disordered dilute media in the nonlinear regime. The approach 
is non-perturbative in the strength of the nonlinearity.
We show that the coherent backscattering enhancement factor
is strongly affected by the nonlinearity, and corroborate these results
by numerical simulations. Our theory can be applied to several
physical scenarios like scattering of light in nonlinear Kerr
media, or propagation of matter waves in disordered potentials.
\end{abstract}

\maketitle

It is already known that the interplay between disorder
and - even very weak - nonlinearity can lead to dramatic changes to the
system's properties: for example, instabilities
occur~\cite{spivak,skipetrov,peter_tobias}, or localization may be
destroyed \cite{shepel}. In the experiments studying the
localization properties of matter waves in speckle potentials
\cite{bec}, the
nonlinear regime, arising from the atomic interactions, is almost
unavoidable. Furthermore, nonlinear behavior is easily observed in coherent
backscattering experiments using cold atoms as scatterers
\cite{thierry}. As a third example, also the random laser exhibits
nonlinearities which potentially influence the structure of
localized laser modes \cite{cao}. In all these
cases, even if the systems are governed by simple
nonlinear wave equations, a precise
description of the impact of this
nonlinearity on the interference effects altering the properties of diffuse wave
propagation is still lacking. Since exact numerical calculations for realistic situations
are at the border of or beyond actual computer capacities, one needs an
efficient theory providing directly disorder averaged quantities. 
For this purpose, the present letter shows that the standard diagrammatic
approach \cite{ishimaru,vanderMark,robert} can be extended
to the nonlinear regime. Using ladder and crossed-like diagrams, we
will derive a nonlinear radiative
transfer equation for the averaged
wave intensity, and then calculate the interference corrections on top
of the nonlinear solution. 

The general frame where our approach can be applied is as follows:
we assume a nonlinear wave equation with 
unique and stationary monochromatic solution. In particular, we assume
that all even orders $\chi^{(n)}$,
$n=2,4,\dots$, of the nonlinear susceptibility vanish, such that the
generation of higher harmonics can be neglected. Furthermore,
the refractive index modifications are small
enough such that we can neglect effects like self-focusing, pattern formation and
solitons \cite{boyd} on the length scale set by the disorder (a mean free path).
Instead, 
the nonlinear effects relevant
for our disordered case can be summarized as follows: firstly, the wave
intensity $I({\bf r})$ becomes a fluctuating quantity, which is
especially important in the nonlinear regime. Secondly,
the usual
picture of weak localization resulting from interference only between pairs of
amplitudes propagating along reversed paths
breaks down in the nonlinear regime. As a consequence of the nonlinear
mixing between different partial waves, weak localization must
rather be
interpreted as a multi-wave interference phenomenon
\cite{wellens05,wellens06b}. In particular,
we will show in the following that the height of the coherent
backscattering peak is strongly
affected by nonlinearities, even if they do respect the reciprocity
symmetry. In contrast to our previous work
\cite{wellens05,wellens06b}, the present approach is
non-perturbative in the strength of the nonlinearity.

At first, we consider an assembly of $N$ point-like
scatterers located at randomly chosen 
positions $\mathbf{r}_i$, $i=1,\dots,N$ inside a sample volume $V$
illuminated by a plane wave $\mathbf{k}_L$, with $|{\bf  k}_L|=k$. We assume the field
radiated by each scatterer to be a nonlinear function $f(E_i)$
of the local field $E_i$.
Since all even orders of
the nonlinear susceptibilities vanish, we can write $f(E)=g(I)E$, where
$I=EE^*$ is the local intensity, and $g(I)$ is proportional to the
polarizability of the scatterers. 
This results in a set of
nonlinear equations for the field at each scatterer:
\begin{equation}
\label{model}
E_i=e^{i\mathbf{k}_L\cdot \mathbf{r}_i}+\sum_{j\neq i}
\frac{e^{ik|{\bf r}_i-{\bf r}_j|}}{4\pi|{\bf r}_i-{\bf r}_j|} g(E_jE_j^*)E_j
\end{equation}
As explained above, we aim at providing a
theory 
providing the
relevant quantities (local intensities, coherent backscattering cone...)
averaged over the
random positions of the scatterers.
In a first step, we will derive an equation for the
mean intensity $\langle I({\bf r})\rangle$. In the dilute regime, where the typical
distances $|{\bf r}_i-{\bf r}_j|$ are 
much larger than the wavelength, we may neglect - in first
approximation - correlations
between the fields emitted by different scatterers. Under this
condition, the scattered field $E_d({\bf r})$ is a
superposition of spherical waves with random relative phases,
depicting thus a speckle pattern. This leads to the well known
Gaussian statistics for the complex field $E_d({\bf r})$ \cite{goodman},
which is thus completely determined by a
single parameter, the mean diffuse intensity $I_d({\bf r})=\langle
|E_d({\bf r})|^2\rangle$. In
addition to the scattered field, there is also a non-fluctuating coherent component
originating directly from the incident field. In total, we have
$E({\bf r})=E_c({\bf r})+E_d({\bf r})$, and the average intensity
splits into a coherent and diffuse part: $\langle I({\bf
  r})\rangle=I_c({\bf r})+I_d({\bf r})$, with $I_c=|E_c|^2$. The mean
density of radiation intensity emitted from point $\mathbf r$ 
is then given by:
\begin{equation} 
K(\mathbf{r}) =  \mathcal{N}\left<ff^*\right>=
\mathcal{N}\left<|g\left(I(\mathbf{r})\right)|^2I(\mathbf{r})\right>\label{k}
\end{equation} 
where $\mathcal{N}=N/V$ denotes the density of scatterers, and
the average $\left<\dots\right>$ is taken over the Gaussian
statistics of the scattered field. 

Between two scattering events, the wave propagates in an effective medium made by
the scatterers. The corresponding refractive index $n$ and 
mean free path $\ell$ are not the same for the
coherent and the diffuse fields, because of their different statistical
properties combined with the nonlinear behavior of the
scatterers. In the dilute regime, the diffuse amplitude can be considered
as a weak probe, such that the complex refraction index reads:
\begin{equation}
n = 1+
\frac{\mathcal{N}}{2k^2}\left<\frac{df}{dE}\right>,\
\frac{1}{\ell}=2k{\rm Im}\{n\} 
\label{index}
\end{equation}
whereas, for the coherent mode, the derivative $d/dE$ is replaced by
$1/E_c$, i.e.
$n_c=1+
\mathcal{N} \left<f\right>/(2k^2E_c)$, and $1/\ell_c=2k{\rm Im}\{n_c\}$. 
Since the results of
the averages depend on $I_c({\bf r})$ and $I_d({\bf r})$, the
nonlinear refractive indices also attain a
spatial dependence $n({\bf r})$ and $n_c({\bf r})$. They describe
average propagation of one
strong and many uncorrelated weak fields. (If there is more than one
strong field, additional phenomena like four-wave mixing occur \cite{boyd}.)

Recollecting all preceding ingredients, the transport equations for
the average intensity read as follows:
\begin{eqnarray}
I_c({\bf r}) & = & e^{-\overline{z/\ell_c}}\label{ic}\\
I_d({\bf r}) &  = & 
\int_V d{\bf r'} \frac{e^{-\overline{|{\bf r}-{\bf
      r'}|/\ell}}}{(4\pi|{\bf r}-{\bf r'}|)^2}K({\bf r'})\label{id}
\end{eqnarray}
Here, $z$ denotes the
distance from the surface of $V$ to $\bf r$, in the direction  
of the incident beam. Furthermore, propagation from ${\bf r'}$ to $\bf
r$ implies a spatial
average of $1/\ell({\bf r})$, which we note as follows:
$\overline{|{\bf r}-{\bf r'}|/\ell}:=|{\bf r}-{\bf r'}|\int_0^1ds/\ell({\bf
      r}-s{\bf r}+s{\bf r'})$, and similarly for the coherent mode
($\overline{z/\ell_c}$).
Since $K$, $\ell$ and $\ell_c$ depend on $I_c({\bf r})$ and $I_d({\bf r})$, the
above Eqs.~(\ref{ic},\ref{id}) form two coupled integral equations,
whose solution we find numerically. Finally, the intensity scattered
into backwards direction (expressed as dimensionless quantity, the
so-called \lq bistatic coefficient\rq \cite{ishimaru}) results as:  
\begin{equation}
\Gamma_L  = \int \frac{d{\bf r}}{4\pi A} e^{-\overline{z/\ell}} K({\bf r})\label{gammal}
\end{equation}
where $A$ denotes the transverse (with respect to the incident beam)
area of the scattering volume $V$.

The validity of the preceding approach has been tested using the
following nonlinear function:
\begin{equation}
g(I)=\frac{2\pi i}{k(1+\alpha I)}\label{modelg}
\end{equation}
which depicts the (elastic) nonlinear behavior of a two-level atom
exposed to an intense laser beam. The nonlinear scatterers described by
Eq.~(\ref{modelg}) are randomly distributed inside a
sphere, with homogeneous density. We
must emphasize that, for this particular model of nonlinearity, the
stationary solution is always found to be unique and stable, as a
consequence of the saturation $g(I)\to 0$
for large $\alpha$. From the numerical solution of Eq.~(\ref{model}), we
calculate the radiated field and intensity outside the cloud in
different directions $\theta$. This procedure is then repeated with many
different configurations giving us the disorder averaged field and
intensity. The results presented in this letter are obtained with 3000
configurations of $1500$ scatterers with density such that
$k\ell=67$ and optical thickness $b=2$ (in the linear limit $\alpha=0$). 
\begin{figure}
\centerline{\includegraphics[height=8cm,angle=-90]{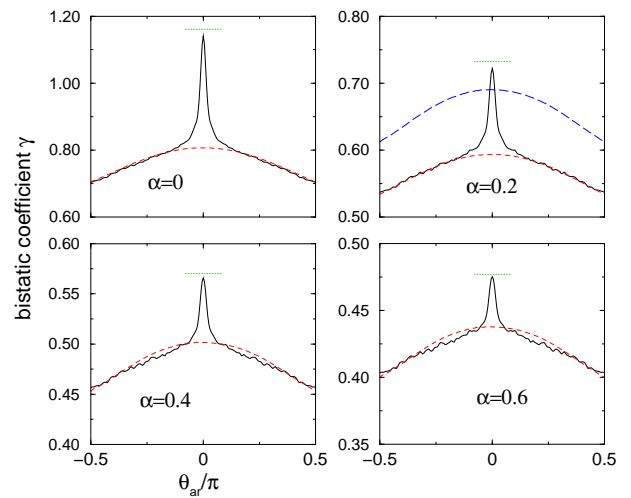}}
\caption{\label{numerics} (Color online) Comparison between exact numerical calculations
  and the theoretical approach (see text for details). The solid line depict
the exact numerical results, whereas the dashed line corresponds
to $\Gamma_L$ including geometrical effects. The dotted line
corresponds to the sum $\Gamma_L+\Gamma_C$ exactly in the backward
direction. The
additional curve (long dashed line) plotted for $\alpha=0.2$ depicts
the results obtained when the fluctuating character of the diffuse
field is not taken into account.}
\end{figure}

The results for the average intensity as a function of the
backscattering angle $\theta$ are depicted in Fig.~\ref{numerics} for
different values of the nonlinear parameter $\alpha=0$, $0.2$, $0.4$ and
$0.6$.  For each plot, the solid line
depicts the exact numerical results, whereas the dashed line corresponds
to $\Gamma_L$, Eq.~(\ref{gammal}), supplemented by a geometrical
factor depending on $\theta$. Away from the backward direction,
the agreement between the exact numerical calculations and our theoretical
prediction for the background is clearly excellent. This is emphasized
by the additional curve (long dashed line) plotted for $\alpha=0.2$
depicting the results obtained when neglecting the
fluctuations of $I(\mathbf{r})$,
for example replacing $\left<|g(I)|^2I\right>$ by
$|g(\left<I\right>)|^2\left<I\right>$ in Eq.~(\ref{k}).

In the backward direction, constructive interference between
reversed scattering paths results in the well-known coherent
backscattering peak. As obvious from Fig.~\ref{numerics}, the
height of this peak is strongly affected by the nonlinearity.
Nevertheless, we are
perfectly able to incorporate these interferences effects in our approach,
see the horizontal dotted lines in Fig.~\ref{numerics}, which depict the
predicted total bistatic coefficient, $\Gamma_L+\Gamma_C$, see
Eq.~\eqref{gammac} below, in the exact backward direction~\cite{fullcone}.
These results are obtained
by a diagrammatic analysis, which we summarize in the following.

\begin{figure}
\centerline{\includegraphics[width=8cm]{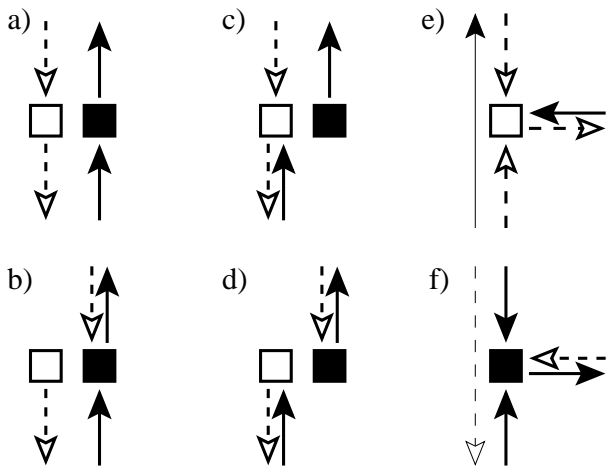}}
\caption{Building blocks for the diagrammatic calculation of nonlinear
  coherent backscattering. Filled
  squares (with outgoing solid arrows) denote the scattered field $f$, and open
  squares (with outgoing dashed arrows) the
  complex conjugate $f^*$. Incoming solid (dashed) arrows represent
  derivatives $d/dE$ ($d/dE^*$). Diagrams (a-d) contribute
  to the cooperon
  cross sections $\kappa$ and $\tilde{\kappa}$, see Eq.~(\ref{kappa}), whereas
  diagrams (e,f) represent nonlinear propagation $\tau^*$ and $\tau$, see Eq.~(\ref{tau}).
\label{diag}}
\end{figure}

As in the linear theory, we calculate the coherent backscattering
effect by so-called \lq crossed\rq\ 
or \lq cooperon\rq\ diagrams \cite{vanderMark}, describing
pairs of reversed scattering paths.
Hence, the individual scatterers are subject to two different incident
probe fields $E$ and $E^*$, which represent the two amplitudes propagating
along the reversed paths. The response
of a scatterer to these two weak probe fields is given by the
derivative $d^2/(dEdE^*)$. Depending on
whether the incident fields act on the dipole $f$ or its complex
conjugate $f^*$, we obtain the four terms represented in
Fig.~\ref{diag}(a-d).
If we denote the sum of diagram (a) + (c) by $\kappa$, and 
(b) + (d) by $\tilde{\kappa}$, the explicit expressions read:
\begin{equation}
\kappa = \mathcal{N}\left<\frac{d}{dE}\left(f\frac{df^*}{dE^*}\right)\right>,\ 
\tilde{\kappa} = \mathcal{N}\left<\frac{d}{dE}\left(f^*\frac{df}{dE^*}\right)\right>\label{kappa}
\end{equation} 
If one of the incident fields originates from the coherent mode, 
$d/dE$ is again replaced by $1/E_c$, i.e. $\kappa_c=\mathcal{N}\left<f df^*/dE^*\right>/E_c$ and
$\tilde{\kappa}_c  =  \mathcal{N}\left<f^*df/dE^*\right>/E_c$. 

Concerning propagation between two scattering events, the refractive index,
Eq.~(\ref{index}), remains unchanged for the
reversed paths. 
In addition to
that, however, we find two other contributions shown in  
Fig.~\ref{diag}(e,f), which exist only as crossed diagrams.
Here, diagram (e) represents the expression:
\begin{equation}
\tau  =
-\frac{i\mathcal{N}}{2k}\left<\frac{d^3f^*}{(dE^*)^2dE}\right>\label{tau}
\end{equation}
and diagram (f) its complex conjugate ($\tau^*$). For the coherent
mode, the above expression is again modified as follows:
$\tau_c = - i{\mathcal N}\left<d^2f^*/(dE^*)^2\right>/(2kE_c)$.
The
propagation of the thin line in Fig.~\ref{diag}(e,f) is unaffected
by the nonlinear event. 
\begin{figure}
\centerline{\includegraphics[width=8cm]{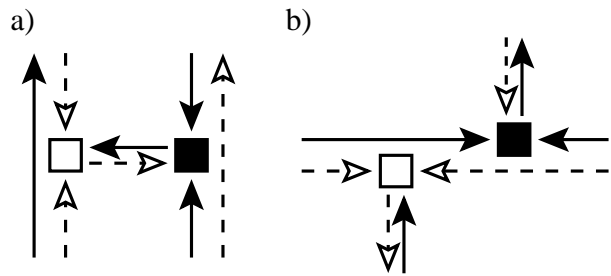}}
\caption{Examples of forbidden (a) and allowed (b) combinations of the
  building blocks shown in Fig.~\ref{diag}. The forbidden combinations
  correspond to those forming a closed loop in the sense that two squares are
  connected by two arrows pointing in
  different directions.\label{forbidden}}
\end{figure}

The crossed transport equation is established by connecting the
building blocks shown in Fig.~\ref{diag} with each other. 
As we have found, however, some
combinations of diagrams represent unphysical 
processes which do not occur in the formal expansion of the
solution of the nonlinear wave equation as a multiple scattering
series. An example is shown in
Fig.~\ref{forbidden}(a). The problem with this diagram is that the
fields radiated by $f^*$ and $f$ 
mutually depend on one another. Therefore, one cannot tell which
one of the two events $f$ or $f^*$ happens before the other one. This
contradicts the multiple scattering series, where the
individual scattering events occur one after the
other. In order to avoid closed loops
like the one shown in Fig.~\ref{forbidden}(a), we must ignore all combinations where
one of the diagrams Fig.~\ref{diag}(c,d) or (e) occurs after
Fig.~\ref{diag}(b,d) or (f) when following the solid arrow along the 
crossed path.

We account for these forbidden diagrams
by splitting the transport equation into two parts,
which we call $C_1$ and $C_2$.
The first part, $C_1$, contains only diagrams Fig.~\ref{diag}(a,c) and (e).
As soon as one of the events Fig.~\ref{diag}(b,d,f) occurs, the crossed
      intensity changes from type $C_1$ to type $C_2$. The
      subsequent propagation of $C_2$ is then given by 
      diagrams Fig.~\ref{diag}(a,b,f). Following these rules, we
      describe the
      propagation of $C_{1,2}$ by transport equations
      similar to Eqs.~(\ref{ic},\ref{id}):
\begin{eqnarray}
\label{set1}
C_c({\bf r}) & = & e^{ik\overline{z(n_c-n^*)}}\label{cc}\\
C_{1}({\bf r}) & = & \int_V d{\bf
  r'}P({\bf r},{\bf r'})\Bigl(\sigma C_{1}+\sigma_c C_c\Bigr)({\bf r'})\label{c1}\\
C_{2}({\bf r}) & = & \int_V d{\bf r'} P({\bf r},{\bf r'})
\Bigl(\sigma^* C_{2}+\tilde\sigma C_{1}+\tilde{\sigma}_cC_c\Bigr)({\bf
  r'})\ \ \ \label{set2}
\end{eqnarray}
where the propagation kernel $P({\bf r},{\bf r'})=\exp(-\overline{|{\bf r}-{\bf
      r'}|/\ell})/(4\pi|{\bf r}-{\bf r'}|)^2$ is the same as in
      Eq.~(\ref{id}), and the cross sections $\sigma$ result 
as follows: $\sigma=\kappa+\ell K \tau$, 
$\tilde{\sigma}=\tilde{\kappa}+\ell K \tau^*\label{sigma}$
and, similarly, $\sigma_c=\kappa_c+\ell K \tau_c$ and
      $\tilde{\sigma}_c=\tilde{\kappa}_c+\ell K \tau_c^*$. 
Finally, the crossed
      bistatic coefficient reads:
\begin{equation}
\Gamma_C  =  \int_V \frac{d{\bf r}}{4\pi A}
e^{ik\overline{z(n-n_c^*)}} \Bigl((\sigma_c^*+\tilde{\sigma}_c^*)C_1+\sigma_c^*
C_2\Bigr)({\bf r})\label{gammac}
\end{equation}
For comparison
      with the background $\Gamma_L$, we define
      diffuson cross sections by writing 
$K=\sigma^{(d)} I_d+\sigma^{(d)}_c I_c$, such that Eq.~(\ref{id})
      attains a form comparable to Eq.~(\ref{c1}).  Exploiting the Gaussian properties
      of the diffuse field, we find: $\sigma^{(d)}={\mathcal
  N}\left<d(ff^*)/dI\right>$
and $\sigma^{(d)}_c={\mathcal
  N}\left<d(ff^*)/dE\right>/E_c^*$. Thus,
the decrease of the
backscattering peak observed in Fig.~\ref{numerics} is traced back to the
      fact that the cooperon
cross sections $\sigma$ and $\tilde{\sigma}$
decrease faster than the diffuson cross section $\sigma^{(d)}$.
Let us note that there also exist other models than
Eq.~(\ref{modelg}), where our theory predicts an increasing coherent backscattering cone.
However, these models might suffer from
speckle instabilities - a point which requires further investigations.

To obtain 
the relatively simple form of
Eqs.~(\ref{set1}-\ref{gammac}), we assume that the scattered intensity
$K({\bf r})$ is approximately constant on length scales comparable to the mean free path
$\ell$, and we neglect some diagrams where the coherent mode $C_c$ is affected by a
nonlinear event $\tau_c$. These approximations are expected to be well fulfilled in
the case of large optical
thickness $b$. In the numerical comparison depicted in
Fig.~\ref{numerics}, where $b$ is not very large, 
we have used the exact version of
Eqs.~(\ref{set1}-\ref{gammac}), which will be published elsewhere.
   
As explained in the introduction, our theoretical scheme also applies
to other types of nonlinear systems, like, for example, the case of linear
scatterers embedded in a homogeneous nonlinear medium:
\begin{equation}
\Delta E({\bf r})+k^2\Bigl(\epsilon({\bf r})+\alpha|E({\bf
  r})|^2\Bigr)E({\bf r})=0\label{hom}
\end{equation}
with $\delta$-correlated disorder $\epsilon({\bf r})$ corresponding to
a (linear) mean free path $\ell_0$.
Here, the dilute medium approximation is valid if
$k\ell_0\gg 1$ and $(\alpha I)^2 k\ell_0\ll 1$. The latter condition
is automatically fulfilled if we assume that we are in the stable
regime, where Eq.~(\ref{hom}) has a unique solution.  According to
\cite{skipetrov}, this is the case (for $\alpha\in{\mathbbm R}$) if
$(\alpha I)^2 b^2(k\ell_0+b)<1$, with $b$ the optical thickness. 

In this case, the diagrammatic method applies in the same way as
described above. In particular, we obtain the following expressions
for the cross sections:
\begin{equation}
\sigma({\bf r})=\sigma_c({\bf r})=\frac{4\pi}{\ell_0}\left[1+ik\ell_0\alpha \left(I_c({\bf
    r})+I_d({\bf r})\right)\right],\label{sigmahom} 
\end{equation}
$\tilde{\sigma}=\tilde{\sigma}_c=-4\pi ik\alpha (I_c+I_d)$,
$\sigma^{(d)}=\sigma^{(d)}_c=4\pi/\ell_0$, and for the mean free paths
$n=\langle\epsilon\rangle+\alpha(I_c+I_d)+i/(2k\ell_0)$ and
$n_c=\langle\epsilon\rangle+\alpha(I_c/2+I_d)+i/(2k\ell_0)$.
In the energy conserving case
$\alpha\in{\mathbbm R}$, it can be shown that $C_2$ does not 
contribute to the real part of the backscattering coefficient $\Gamma_C$.
Then, it follows from Eqs.~(\ref{c1}) and (\ref{sigmahom}), that the
nonlinearity introduces  
a phase difference $\Delta\phi=Mk\ell_0\alpha
I$ between reversed paths undergoing $M$ linear scattering events. Since
$\langle M\rangle\propto b$, we predict a significant reduction of the
coherent backscattering peak if 
$b k\ell_0\alpha I\simeq 1$ (which is still inside the stable regime if
$k\ell_0$ is large).

In summary, we have extended the usual diagrammatic approach
to take into account nonlinear effects for the coherent transport in
disordered systems beyond the perturbative regime. The excellent
agreement with direct numerical simulations emphasizes the validity of our
approach. It readily
applies for many different nonlinear wave equations. Eq.~(\ref{hom}),
for example, is mathematically equivalent to the
Gross-Pitaeskii equation describing nonlinear propagation of
matter waves in random potentials. In the latter case, this method will
allow us to describe 
the localization
properties of the mean field. Extending the present
approach within the Bogolioubov framework, it will be possible to
understand how these localization properties are affected by the
non-condensed fraction of the atoms. 

We thank D.~Delande and C.~Miniatura for fruitful
discussions. T.W. acknowledges support from the DFG.


\begin{thebibliography}{99}

\bibitem{spivak}B. Spivak and A. Zyuzin, Phys. Rev. Lett. {\bf
    84}, 1970 (2000) 
\bibitem{skipetrov}S.E. Skipetrov and R. Maynard, Phys. Rev. Lett. {\bf 85}, 736 (2000) 
\bibitem{peter_tobias}T. Paul \textit{et al.}, Phys. Rev. A
  \textbf{72}, 063621 (2005)
\bibitem{shepel}D. L. Shepelyansky, Phys. Rev. Lett {\bf 70}, 1787
  (1993)
\bibitem{bec}D. Cl\'ement \textit{et al.}, Phys. Rev. Lett {\bf 95}
  170409 (2005); C. Fort \textit{et al.}, Phys. Rev. Lett. {\bf 95}
  170410 (2005); T. Schulte \textit{et al.}, Phys. Rev. Lett. {\bf 95}
  170411 (2005); L. Sanchez-Palencia  \textit{et al.},
  Phys. Rev. Lett. \textbf{98}, 210401 (2007) 
\bibitem{thierry} T. Chaneli\`ere \textit{et al.}, Phys. Rev. E
  \textbf{70} 036602 (2004)
\bibitem{cao}H. Cao, Waves Ranom Media {\bf 13} R1 (2003)
\bibitem{ishimaru}A. Ishimaru,{\em Wave Propagation and Scattering in Random
    Media} (Academic, New York, 1978), Vols. I and II
\bibitem{vanderMark} M. B. van der Mark, M. P. van Albada, and
  A. Lagendijk, Phys. Rev. B {\bf 37}, 3575 (1988)
\bibitem{robert}R. C. Kuhn \textit{et al.},
  Phys. Rev. Lett. \textbf{95}, 250403 (2005)
\bibitem{boyd}R. W. Boyd, {\em Nonlinear Optics} (Academic, San Diego, 1992).
\bibitem{wellens05} T. Wellens, B. Gr{\'e}maud, D. Delande, and C.
  Miniatura, Phys. Rev. E {\bf 71} 055603(R) (2005); Phys. Rev. A {\bf 73} 013802 (2006)
\bibitem{wellens06b} T. Wellens and B. Gr{\'e}maud,
J. Phys. B: At. Mol. Opt. Phys. {\bf 39}, 4719 (2006)
\bibitem{goodman}J. W. Goodman, J. Opt. Soc. Am. {\bf 66}, 1145 (1976)
\bibitem{fullcone} Our theory predicts the full shape of the cone, but
  if $\theta>0$ the invariance by rotation is lost for the crossed
intensities $C_{c,1,2}$, leading to much larger linear systems.

\end{thebibliography}
\end{document}